# Super-Resolution Optical Coherence Tomography Using Diffusion Model-Based Plug-and-Play Priors

Yaning Wang[1,*,†], Jinglun Yu[1,*,†], Wenhan Guo[2], Yu Sun[1], Jin U. Kang[1]
[1]Whiting School of Engineering, Johns Hopkins University, Baltimore, MD, USA
[2]Physics and Astronomy Department, Pomona College, Claremont, CA, USA
{ywang511, jyun146, ysun214, jkang}@jh.edu, wgaa2021@mymail.pomona.edu
[*]Equal contribution    [†]Corresponding author

## Abstract

We propose an OCT super-resolution framework based on a plug-and-play diffusion model (PnP-DM) to reconstruct high-quality images from sparse measurements (OCT B-mode corneal images). Our method formulates reconstruction as an inverse problem, combining a diffusion prior with Markov chain Monte Carlo sampling for efficient posterior inference. We collect high-speed under-sampled B-mode corneal images and apply a deep learning-based up-sampling pipeline to build realistic training pairs. Evaluations on *in vivo* and *ex vivo* fish-eye corneal models show that PnP-DM outperforms conventional 2D-UNet baselines, producing sharper structures and better noise suppression. This approach advances high-fidelity OCT imaging in high-speed acquisition for clinical applications.

## 1 Introduction

Advancing beyond diagnostic imaging, OCT is increasingly being used for intraoperative imaing to visualize fine tissue microstructures and guide microsurgeries in real time [1-6]. Among all application areas, ophthalmology remains the primary field where OCT has been playing a crucial role.

However, one of the main challenges in ophthalmic OCT imaging is the presence of involuntary eye motion during image acquisition [7]. The motions range from fixational movements such as tremors, drifts, and micro saccades, to the axial displacements induced by vascular pulsation and respiration, which can distort the acquired OCT voxels and limit the reproducibility of quantitative measurements [8]. Such distortions are especially problematic in clinics, where patients' motions may be amplified, which degrades the overall image quality and affects postprocessing analysis.

To address these challenges, both hardware- and software-based motion correction methods have been explored. Hardware solutions, which typically involve auxiliary sensors to track ocular movement, often require complex and expensive system designs [9, 10]. On the other hand, software-based approaches rely on hand-crafted features or model-driven algorithms [11], which can be limited in accuracy, especially under unpredictable practice in real world.

In recent years, deep learning (DL) techniques have shown promise in enhancing OCT image quality through tasks like image motion-deblur or super-resolution (ISR) [12, 13]. These approaches aim to reconstruct high-resolution (HR) images from low-resolution (LR) observations affected by noise, blur, and motion artifacts. However, existing models often oversimplify ocular motion as mere rigid translation, neglecting complex non-rigid deformations such as rotational and tissue-based dynamics. Furthermore, many supervised DL models trained on synthetic LR-HR pairs with pixel-wise losses struggle to generalize to real-world degradation, resulting in over-smoothed reconstructions lacking structural fidelity [14-16].

To improve the perceptual realism and robustness of real-world ISR, recent works have incorporated natural image priors using generative adversarial networks (GANs). While GAN-based methods have demonstrated encouraging results—particularly in domains like face restoration [17, 18]—they often suffer from training instability and struggle to model the wide diversity of natural scenes. Some alternative approaches leverage sparsity priors through residual networks and compressed sensing, which can capture intrinsic image patterns more effectively [19, 20].

A more recent and promising direction involves the use of generative diffusion models (DMs) [21]. These models, particularly large-scale pretrained architectures [22], have demonstrated superior generative capabilities across a range of downstream tasks. Their ability to encode powerful natural image priors makes them especially suitable for enhancing the realism and structural integrity of real-world ISR outputs.

In this work, we propose interative plug-and-play framework for robust OCT image reconstruction during sparse data acquisition that incorporates a novel plug-and-play diffusion model (PnP-DM). Our method proposed motion-compensated OCT B-scan corneal images towards clinical settings. The framework is formulated as an image inverse problem, where we aim to reconstruct high-resolution OCT volumes from sparse and noisy measurements. To build realistic LR-HR training pairs, we collect undersampled high-speed scans and apply a DL-based up sampling pipeline. Our PnP-DM algorithm integrates diffusion priors within a Markov chain Monte Carlo (MCMC) sampling framework, allowing for efficient posterior sampling and high-fidelity reconstruction. We evaluate our approach on both *in vivo* and *ex vivo* fish-eye models and demonstrate that our method outperforms existing 2D-UNet architecture for high-resolution OCT imaging recovery.

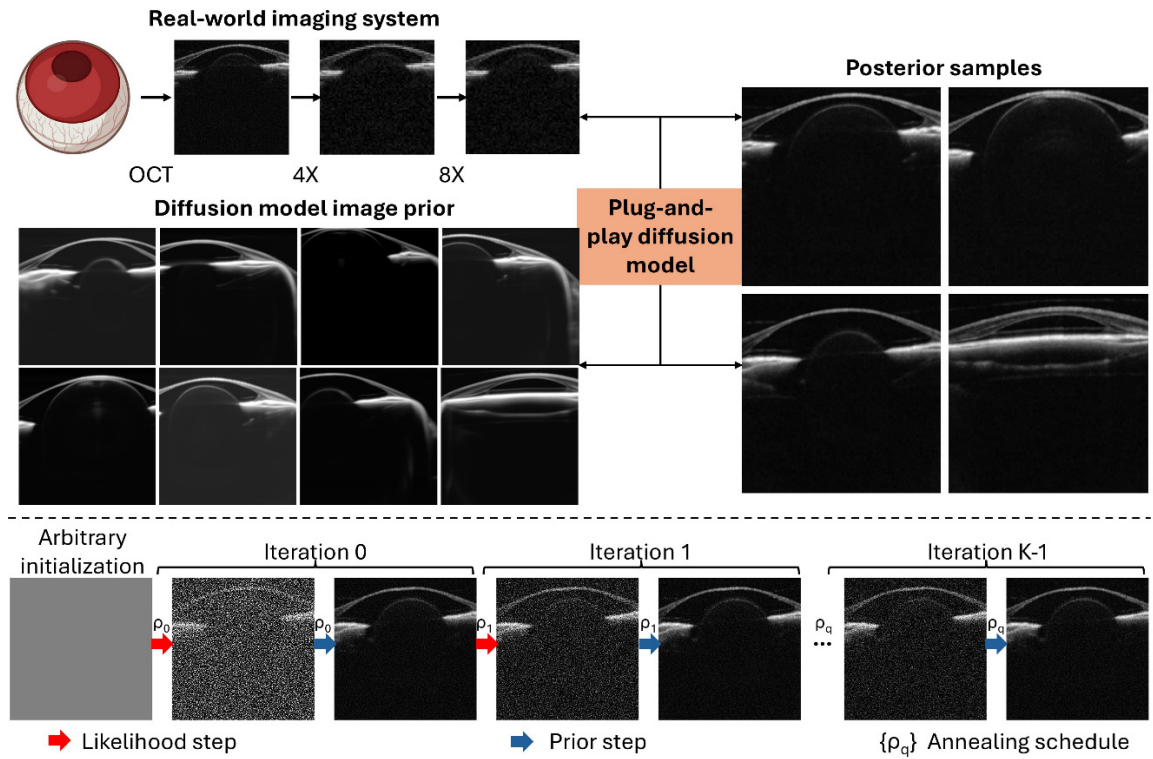

Fig. 1: Overview of the proposed PnP-DM method for posterior sampling using *in vivo* fisheye corneal OCT B-scans. The framework integrates real-world measurements with a diffusion model prior trained on the same dataset, alternating between data-consistent likelihood updates and EDM-based prior refinement.

## 2 Method

This section presents the PnP-DM approach proposed by Wu et al. [23]. for OCT super-resolution, an iterative plug-and-play framework designed to generate high-fidelity outputs across a continuous spectrum of image resolutions, enabling super-resolution reconstruction adaptable to various undersampling rates. The overall architecture of PnP-DM is illustrated in Fig. 1.

## 2.1 Problem Modeling

The goal of OCT super-resolution is to estimate a HR OCT image $x \epsilon R^n$ from a given LR observation $y \epsilon R^m$. The measurement process is modeled as,

$$y = A(x) + n, \tag{1}$$

where $A(\cdot): R^n \to R^m$ denotes the forward linear operator, and $n \epsilon R^m$ represents random measurement noise. Specifically, for our proposed super-resolution task, the forward model is defined as,

$$y \sim N(P_f x, \sigma_y^2 I), \tag{2}$$

where $P_f \epsilon R^{n_f \times n}$ implements a block averaging filter that downsamples the image by a factor of $f$. In our setup, we set $f = 4$ (from 256×256 to 64×64), and the singular value decomposition (SVD) of $P_f$ was computed following the implementation of [24].

To model the complex natural image prior, we employ diffusion models. Specifically, we adopt a PnP-DM framework that leverages DMs as implicit priors for posterior sampling (illustrated in Fig. 1). The method alternates between a likelihood step, enforcing data consistency, and a prior step, performing denoising posterior sampling via the Split Gibbs Sampler (SGS) [25].

## 2.2 Likelihood Step

At iteration $q$, the likelihood step samples are given as,

$$z_{(q)} \sim \pi^{Z|X=x^{(q)}}(z) \propto exp\left(-f(z;y) - \frac{1}{2\rho^2} \| x^{(q)} - z \|_2^2 \right), \tag{3}$$

where $f(z;y)$ is the data-fidelity term. In our case, since the forward model $A$ is linear and the noise $n \sim N(0, \Sigma)$ is zero-mean Gaussian, the potential function becomes:

$$f(x;y) = \frac{1}{2} \| y - Ax \|_\Sigma^2, \tag{4}$$

with the weighted norm defined as $\| v \|_\Sigma^2 = \langle v, \Sigma^{-1} v \rangle$.

Thus, the conditional distribution followed by Eq. 3 for the likelihood step is given by,

$$\pi^{Z|X=x} = N(m(x), \Lambda^{-1}), \tag{5}$$

where,

$$\Lambda = A^T \Sigma^{-1} A + \frac{1}{\rho^2} I, m(x) = \Lambda^{-1}\left(A^T \Sigma^{-1} y + \frac{1}{\rho^2} x\right). \tag{6}$$

The SVD of $A$ is utilized to efficiently compute matrix operations and significantly reduce computational costs.

## 2.3 Prior Step: EDM Denoising

The prior step addresses the denoising posterior sampling problem using diffusion models. We adopt the EDM framework [26], which unifies various diffusion formulations for unconditional image generation. The reverse-time stochastic differential equation (SDE) in EDM is given by,

$$dx_t = \left(\frac{\dot{s}(t)}{s(t)} x_t - 2s(t)^2 \dot{\sigma}(t)\sigma(t) \nabla log p\left(\frac{x_t}{s(t)}; \sigma(t)\right)\right) dt + s(t)\sqrt{2\dot{\sigma}(t)\sigma(t)} d\overline{w_t}, \tag{7}$$

where $d\overline{w_t}$ is an n-dimensional Wiener process running backward in time $t$, $\sigma(t) > 0$ is the noise schedule, and $s(t)$ is the scaling schedule. The key property is that for any $t$, $x_t/s(t) \sim p(x; \sigma(t))$. Solving this SDE backward enables us to transition from noisy samples to the clean image distribution at $t = 0$. In our experiments, we used the original EDM training setup as proposed by Song et al. [27].

## 2.4 Putting It All Together

The complete PnP-DM algorithm alternates between the likelihood and prior steps with an annealing schedule $\{\rho_q\}$ for the coupling parameter $\rho$.

We adopt an exponential decay schedule for $\rho$, which accelerates mixing and avoids getting trapped in local minima in highly ill-posed inverse problems. The schedule is defined as:
- Initial $\rho_0 = 10$,

- Minimum $\rho_{min} = 0.3$,
- Decay rate $\alpha = 0.9$.

## 3 Experiments

To demonstrate the effectiveness of our PnP-DM based framework in enhancing OCT image resolution from sparse measurements, we conduct extensive experiments on both in vivo and ex vivo fisheye corneal datasets. Our evaluations focus on reconstruction fidelity, structural preservation, and noise suppression under clinically relevant undersampling conditions. We compare our method against classical interpolation, supervised deep learning models, and several diffusion-based priors, highlighting the strengths and robustness of our plug-and-play approach across various imaging scenarios.

### 3.1 Data Preparation

The training datasets consist of 164 volumetric OCT scans of *ex vivo* fisheyes, each with a resolution of 1024 × 1024 × 128 voxels. All volumes were normalized to an intensity range of [0, 1], were grayscale images. For multi-resolution training, the high-resolution images (N = 20,992) were down sampled to 256 × 256 resolution by sparsely sampling along the depth and fast scanning axes with a step size of 4. For validation, volumetric datasets were collected from live fisheyes. Model performance was evaluated by reconstructing under sampled test B-scans and comparing the outputs to their corresponding high-resolution images acquired under conventional scanning settings.

### 3.2 Implementation Details

All experiments were conducted on NVIDIA A100 GPUs. The proposed method was implemented using the PyTorch framework [29]. The network was trained for 2,000 epochs with a learning rate of 1e-5 and a decay factor of 0.9, using the Adam optimizer [29]. Training the network with 256 × 256 input images took approximately 26 hours. A dropout rate of 0.05 was applied during training. Since the pre-trained score function was trained with the VP-SDE formulation [21], we convert it to the EDM formulation by applying the VP preconditioning [26]. For testing with PnP-DM models, each image was processed over 100 iterations to ensure convergence, with each Markov chain taking around 80 seconds. All the network parameters are trained with the default FloatPoint-16bit (FP16) datatype.

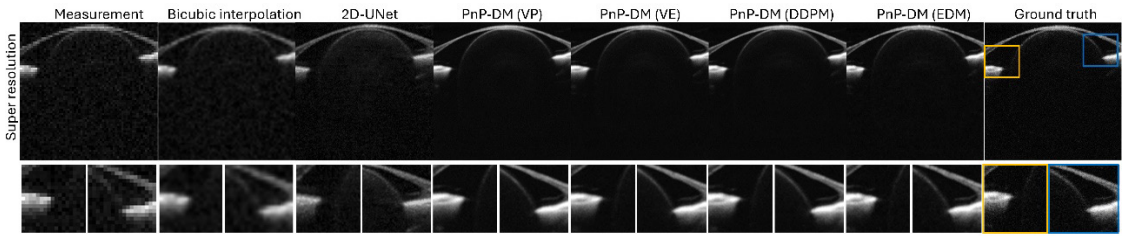

Fig. 2 : Visual examples for the super resolution problem.

### 3.3 Compared Methods

To evaluate the performance of our proposed deep learning network, we compare our methods with several baseline methods as below:
- Bicubic interpolation, a classical non-learning-based method.
- 2D-UNet, a widely-used architecture for medical image super-resolution.
- PnP-DM with different diffusion priors, including VP-SDE (VP), VE-SDE (VE) [27], iDDPM [21], and EDM [26].

For evaluation, we employ three metrics: peak signal-to-noise ratio (PSNR), structural similarity index measure (SSIM), and root mean square error (RMSE). For each diffusion-based method, we draw 100 random samples, compute their means, and report the quantitative metrics on the mean output.

### 3.4 Qualitative Comparisons

Fig. 2 presents qualitative comparisons with the state-of-the-art DL-based method on the 64×64→256×256 OCT super-resolution task. Although bicubic interpolation enables fast reconstructions, it often results in outputs that lack high frequency details and introduce noticeable noise, especially under low sampling conditions where it fails to effectively suppress speckle noise. In contrast, our proposed method produces sharper, more consistent textures that better match the ground truth. The 2D-UNet baseline significantly blurs fine structures and tissue details (e.g., the iris boundary), while our PnP-DM reconstructions better preserve anatomical boundaries, such as the edges of crystalline lens. From a perceptual standpoint, there is little noticeable difference between the reconstructions using different diffusion priors within the PnP-DM framework—all variants can generate HR images that appear both natural and structurally accurate.

Table 1: Quantitative evaluation on super-resolution OCT for 100 fisheye corneal OCT images, **Bold**: best.

|        | BI     | 2D-UNet | PnP-DM (VP) | PnP-DM (VE) | PnP-DM (DDPM) | PnP-DM (EDM) |
|--------|--------|---------|-------------|-------------|---------------|--------------|
| PSNR↑  | 28.13  | 30.07   | 30.48       | **32.50**   | 32.37         | 32.14        |
| SSIM↑  | 0.4607 | 0.4863  | 0.7163      | **0.7224**  | 0.7214        | 0.7150       |
| LPIPS↓ | 0.4024 | 0.2072  | 0.1415      | 0.1257      | 0.1264        | **0.1201**   |

### 3.5 Quantitative Comparisons

Table 1 reports the quantitative results across different methods. Our PnP-DM framework consistently outperforms the baseline models in terms of PSNR, SSIM, and RMSE. Among the diffusion prior variants, VE and EDM show slightly better performance compared to VP and DDPM, particularly under our proposed linear forward models relevant to OCT super-resolution.

### 3.6 Ablation studies

The Effectiveness of the Plug-in Priors to validate the importance of plug-in priors, we conducted an ablation study by replacing the EDM model (trained on OCT fish cornea datasets) with a model trained on a grayscale version of the FFHQ human face dataset. Qualitative results are shown in Fig. 3.

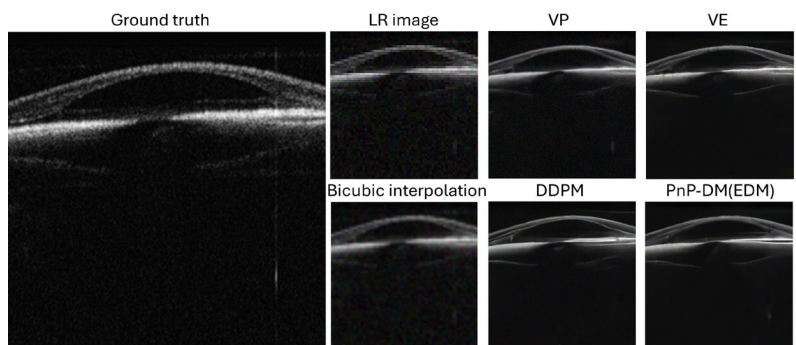

Fig. 3: Visual examples for super-resolution task using human face priors with EDM model.

We observe that the face-trained score functions fail to reconstruct the meaningful corneal structures and exhibit a little mismatch with the measurements, generating fake corneal membrane in the posterior stroma, especially for the DDPM and EDM variants. Despite this, the diffusion models still outperform conventional baselines in speckle noise removal, highlighting the robustness of diffusion-based priors even when they are not perfectly aligned with the target domain.

## 4 Conclusion

We introduced a deep learning framework for super-resolution OCT imaging, leveraging a novel PnP-DM for high-quality reconstruction from sparse measurements. Our method models OCT imaging as an inverse problem and combines diffusion priors with MCMC sampling for efficient and robust inference.

Through experiments on *in vivo* and *ex vivo* fish-eye datasets, PnP-DM demonstrates superior performance over conventional UNet-based methods, achieving sharper reconstructions and better noise suppression. This work highlights the potential of diffusion-driven approaches for advancing high-speed, high-resolution OCT imaging in clinical applications.

### 4.1 Limitations

Our current model processes the entire image simultaneously during both the likelihood and prior steps, which may lead to computational challenges for motion compensation and large-scale volumetric OCT reconstruction.

**Acknowledgments**

We thank Dr. Ruizhi Zuo for his assistance with data collection.